# QRMW: Quantum representation of multi wavelength images

Engin ŞAHİN[1,*], İhsan YILMAZ[2]

[1] Department of Computer and Instructional Technologies Education, Faculty of Education, Çanakkale Onsekiz Mart University, Çanakkale, Turkey

[2] Department of Computer Engineering, Faculty of Engineering, Çanakkale Onsekiz Mart University, Çanakkale, Turkey

*Correspondence: enginsahin@comu.edu.tr

**Abstract:** In this study, we propose quantum representation of multi wavelength images (QRMW) which gives preparation and retrieving procedures of quantum images. Proposed QRMW model represents multi-channel and $2^n \times 2^m$ images. Also, we present image comparison and some image operations based on QRMW model. Comparing our model with the models in literature, QRMW model has less time complexity. Also QRMW model uses fewer qubits than existing models in the literature.

1.   Introduction

There are recently many developments in the quantum computing due to the advantages according to classical computing. One of them is quantum digital image processing. Digital image processing is an important field of technologies such as defense, medical and imaging devices etc. Considering the developments in today's imaging devices and their needs in this direction, the memory required for the images increases when size of the images grows. The operations performed on the images with the classic calculations take a very long time even if very powerful computers are used. Therefore, it is necessary to find high performance methods to store and process images. These operations can be done in a very short time with the parallel processing feature of quantum-based computations. A quantum representation of the image and algorithms based on the representation models are revealed for quantum image processing, respectively.

Various quantum image representation models are proposed to store and process image information. Venegas-Andraca [1] revealed idea of Qubit Lattice which is a representation of quantum image. Venegas-Andraca and Ball [2] proposed the Entangled Image model which uses entangled qubits. Latorre [3] presented Real Ket model through image compression in quantum context. Le et al. [4] presented a flexible representation of quantum images (FRQI) model for quantum computers. In their study, color information is kept for display of images in the form of a normalized state that collecting information about colors and their locations. Zhang et al. [5] presented a novel enhanced quantum representation of digital images (NEQR) model which uses the basis states of a qubit sequence to store the grayscale values at each pixel instead of the probability amplitude of a qubit. Zhang et al. [6] presented quantum image representation for log-polar images (QUALPI) model. Sun et al. [7] proposed the RGB three-channel representation of quantum images (MCQI) model by developing the FRQI model. Yuan et al. [8] presented a simple quantum representation of infrared images (SQR) model by using Qubit Lattice and FRQI methods. Abdolmaleky et al. [9] introduced the RGB three-channel representation for quantum colored images (QMCR) model by using the NEQR model.

At most of the studies in this field, quantum image representations have been presented for single or triple color (RGB) images. It well known that image processing is important at different wavelength in today's technologies. In this respect, there are a wide range of applications employing hyperspectral imaging, ranging from satellite based/airborne remote sensing and military target detection to industrial quality control and lab applications in medicine and biophysics. Therefore, it is important to create a representation model of quantum image that will include all wavelength. There is no representation model for multi-wavelength images in the literature. If the exiting RGB models QMCR and MCQI are developed for more than three channels, they will use too many qubits and will have more time complexity.

In this paper, a quantum representation of multi-wavelength images (QRMW) model is proposed for images that can be formed for all wavelength of light. The proposed QRMW model uses fewer qubits and has less time complexity than existing models in the literature. These advantages are explained in the following sections with the comparisons. This paper is organized as follows: In section 2, we introduce latest related works which is used in comparison with our model. In section 3, we describe the newly proposed QRMW image representation model which gives the procedures of preparation and retrieving of quantum image. In section 4, image compression and some image operations based on QRMW are presented. Finally, conclusion section is given.

## 2. Related Works

In classical computers, image pixels are stored in the classical bits through the photoelectric conversion system. In quantum computers, the storage unit is qubit, so images must be stored in qubit. So far, proposed quantum image representation models [1-9] represent $\mathbf{2^n \times 2^n}$ images with single or triple channels. Image retrieving is probabilistic in the imaging models that hold color information at amplitude (for example, the models based on FRQI). However, retrieving is non-probabilistic in the imaging models where the color information is held at basis states of the qubit sequence (for example, models based on NEQR).

## 2.1. The MCQI model

The MCQI model was originally proposed by Sun et al. [7]. This model stores information about the colors and the position of each pixel of image at the amplitudes. This model for $2^n \times 2^n$ image can be mathematically expressed as follows [7]:

$$|I\rangle = \frac{1}{2^{n+1}} \sum_{i=0}^{2^{2n}-1} |C^i_{RGB\alpha}\rangle \otimes |i\rangle$$

$$\begin{aligned}|C^i_{RGB\alpha}\rangle = &\cos\theta^i_R |000\rangle + \cos\theta^i_G |001\rangle + \cos\theta^i_B |010\rangle + \cos\theta^i_\alpha |011\rangle + \\ &+ \sin\theta^i_R |100\rangle + \sin\theta^i_G |101\rangle + \sin\theta^i_B |110\rangle + \sin\theta^i_\alpha |111\rangle\end{aligned}$$

Where $|000\rangle, |001\rangle \ldots |111\rangle$ are the computational basis states of three qubits, $\theta^i_R, \theta^i_G, \theta^i_B, \theta^i_\alpha$ are four angels which are used to encode the information of color channels of the i-th pixel, $|i\rangle$ for i = 0, 1, . . ., $2^{2n}-1$ are $2^{2n}-$D computational basis states [7].

## 2.2. The QMCR model

The QMCR model was originally proposed by Abdolmalekya et al. [9]. This model is based on the NEQR. This model for $2^n \times 2^n$ image can be mathematically expressed as follows [9]:

$$|I\rangle = \frac{1}{2^n} \sum_{y=0}^{2^n-1} \sum_{x=0}^{2^n-1} |C_{RGByx}\rangle \otimes |yx\rangle$$

Where the state $|C_{RGByx}\rangle$ is used to encode the information of the R, G and B channels ($2^q$ gray range of each channel) of the *yx* pixel. The state $|C_{RGByx}\rangle$ is defined as follows:

$$|C_{RGByx}\rangle = |R_{yx}\rangle|G_{yx}\rangle|B_{yx}\rangle$$

$$|R_{yx}\rangle = |r^{q-1}_{yx} r^{q-2}_{yx} \ldots r^0_{yx}\rangle, |G_{yx}\rangle = |g^{q-1}_{yx} g^{q-2}_{yx} \ldots g^0_{yx}\rangle, |B_{yx}\rangle = |b^{q-1}_{yx} b^{q-2}_{yx} \ldots b^0_{yx}\rangle$$

where $r^k_{yx}, g^k_{yx}, b^k_{yx} \in \{0,1\}$ and $R_{yx}, G_{yx}, B_{yx} \in \{0,1,\ldots, 2^q-1\}$ [9].

## 3. QRMW model

It is important to set an image model with multi-channel. In this paper, we propose a QRMW model with multi-channel (more than three) and $2^n \times 2^m$ images. In this section, the preparation and presentation of the image in the QRMW model will be explained in details.

This new model uses the basis states of qubit sequence to store the values at different wavelength of each pixel of the image. QRMW uses three separate register qubit sequences to hold the position, wavelength, and color values of each pixel. QRMW keeps the whole image in the superposition of the three qubit sequences. The first qubit sequence is used to encode the color values corresponding to the respective wavelength channel of the pixels in the image, the second qubit sequence is used to encode the wavelength channel information and the third qubit sequence is used to encode position information. Suppose that the number of wavelength channel is *cn* and the color-scale in the channels is $2^q$ for the image. Where, $2^q$ represents the maximum value at any wavelength. In our model, *b*-qubits ($b = ceil(\log_2 cn)$) for *cn*-wavelength, q-qubits for color scale and *(n+m)*-qubits for position information are required for $2^n \times 2^m$ image. Color value (with $2^q$ scale) for the channel of the (*y,x*) position of the image can be expressed in binary string as follows

$$f(\lambda, y, x) = c_{\lambda yx}^0 c_{\lambda yx}^1 \dots c_{\lambda yx}^{q-2} c_{\lambda yx}^{q-1}, \quad c_{\lambda yx}^k \in [0,1], \quad f(\lambda, y, x) \in [0, 2^q - 1] \quad (1)$$

The representative expression of a quantum image for a $2^n \times 2^m$ QRMW image can be written as

$$|I\rangle = \frac{1}{\sqrt{2^{b+n+m}}} \sum_{\lambda=0}^{2^b-1} \sum_{y=0}^{2^n-1} \sum_{x=0}^{2^m-1} |f(\lambda, y, x)\rangle \otimes |\lambda\rangle \otimes |yx\rangle \quad (2)$$

where $\lambda$ is a channel information and $yx$ is a position information. The rest of the paper we used $\lambda$ for the channel information and $yx$ for the position information. 2×2 QRMW sample image with [0-255] color scale and 4-channels is shown in Figure 1. In this sample, 12 qubits are used for color, channel and position information. The channel and position information are stored in the superposition of second and third register qubits, respectively.

### 3.1. Preparing procedure

In this section, the preparation process for QRMW images will be explained. In this model, QRMW image is prepared in two steps.

**Step 1:** In the first step of the preparing procedure, $q+b+n+m$ qubits should be prepared and these qubits should be set to $|0\rangle$ as the initial state. The initial state can be expressed as

$$|\psi_0\rangle = |0\rangle^{\otimes q+b+n+m} \tag{3}$$

The identity (I) and Hadamard (H) transforms are applied to convert the initial state $|\psi_0\rangle$ to the intermediate state $|\psi_1\rangle$. We define all quantum operations in the first step as $U_1$

$$U_1 = I^{\otimes q} \otimes H^{\otimes b+n+m} \tag{4}$$

where $I = \begin{bmatrix} 1 & 0 \\ 0 & 1 \end{bmatrix}$, $H = \frac{1}{\sqrt{2}}\begin{bmatrix} 1 & 1 \\ 1 & -1 \end{bmatrix}$, [8]. By applying $U_1$ transformation operator on the state $|\psi_0\rangle$, intermediate state $|\psi_1\rangle$ is obtained as follows.

$$U_1(|\psi_0\rangle) = |\psi_1\rangle = (I|0\rangle)^{\otimes q} \otimes (H|0\rangle)^{\otimes b+n+m}$$
$$= \frac{1}{\sqrt{2^{b+n+m}}} \sum_{\lambda=0}^{2^b-1} \sum_{y=0}^{2^n-1} \sum_{x=0}^{2^m-1} |0\rangle^{\otimes q} \otimes |\lambda y x\rangle \tag{5}$$

The quantum circuit of the first step is given in Figure 2a. After this step, the position and channel information of all the pixels will be stored in the QRMW quantum model. All of the pixels are stored in the superposition of the qubit sequence at the intermediate state $|\psi_1\rangle$.

**Step 2:** In order to set the color information of each channel of each pixel, the second step consists of $2^{b+n+m}$ sub-processes. The quantum sub-process $U_{\lambda yx}$ of the channel $(\lambda)$ of the pixel $(y, x)$ is as follows

$$U_{\lambda yx} = I^{\otimes q} \otimes \sum_{k=0}^{2^b-1} \sum_{j=0}^{2^n-1} \sum_{i=0}^{2^m-1} |kji\rangle\langle kji| + \Omega_{\lambda yx}, \; for \; kji \neq \lambda yx \qquad (6)$$

Where the quantum operator $\Omega_{\lambda yx}$ which sets the color value of the channel $(\lambda)$ of the corresponding pixel $(y, x)$ is

$$\Omega_{\lambda yx} = |f(\lambda, y, x)\rangle|\lambda yx\rangle\langle 0|^{\otimes q}\langle \lambda yx| \qquad (7)$$

The quantum circuit of the $\Omega_{\lambda yx}$ operator is given in Figure 2b.

After applying $U_{\lambda yx}$, the intermediate state $|\psi_1\rangle$ turns to Eq. (8).

$$\begin{aligned} U_{\lambda yx}(|\psi_1\rangle) &= U_{\lambda yx}\left(\frac{1}{\sqrt{2^{b+n+m}}} \sum_{k=0}^{2^b-1} \sum_{j=0}^{2^n-1} \sum_{i=0}^{2^m-1} |0\rangle^{\otimes q} |kji\rangle\right) \\ &= \frac{1}{\sqrt{2^{b+n+m}}}\left(\sum_{k=0}^{2^b-1} \sum_{j=0}^{2^n-1} \sum_{i=0,kji\neq\lambda yx}^{2^m-1} |0\rangle^{\otimes q} |kji\rangle + |f(\lambda, y, x)\rangle|\lambda yx\rangle\right) \end{aligned} \qquad (8)$$

Where $\langle 0^{\otimes q}\lambda yx|0^{\otimes q}\lambda yx\rangle = 1$ due to orthogonality. $U_{\lambda yx}$ sub-operation only sets the color value of the corresponding channel of the pixel in the corresponding position. The operator $U_2$ is defined as follows

$$U_2 = \prod_{\lambda=0}^{2^b-1} \prod_{y=0}^{2^n-1} \prod_{x=0}^{2^m-1} U_{\lambda yx} \qquad (9)$$

The operator $U_2$ includes all the sub-operations that set the color values of each channel of each pixel. After the application of the $U_2$ quantum operator to the $|\psi_1\rangle$ intermediate state, the final state $|\psi_2\rangle$ which represents the QRMW quantum image is obtained as follows.

$$|\psi_2\rangle = U_2(|\psi_1\rangle) = \frac{1}{\sqrt{2^{b+n+m}}} \left( \sum_{\lambda=0}^{2^b-1} \sum_{y=0}^{2^n-1} \sum_{x=0}^{2^m-1} |f(\lambda, y, x)\rangle |\lambda y x\rangle \right) \qquad (10)$$

After these two steps, the preparation procedure is over. The time complexity of the quantum operator $U_1$ in step 1 is $O(q + b + n + m)$. The time complexities of the quantum operators $U_{\lambda yx}$, $\Omega_{\lambda yx}$ and $U_2$ in step2 are respectively $O(q)$, $O(1)$, $O(q \times 2^{b+n+m})$. Let us compare qubit numbers and time complexity of our QRMW model with the models in the literature.

Numbers of qubits of the QRMW, MCQI and QMCR with 4-channels and $2^q$ color scale for $2^n \times 2^n$ images are shown in Table 1.

It is seen that the number of qubits used in MCQI model is less than QRMW model for fewer channels. However, as the number of channels increases, the number of qubits used in QRMW model is less than MCQI model. Also, it is clear that the QRMW model uses fewer qubits than the QMCR model for all channel. The difference between QRMW and QMCR will increase as the number of channels increases.

Storing of position information is the same in the all models. Storing of color information is different in the all models.

In the QRMW model, $2^n \times 2^m$ rectangular images and $2^n \times 2^n$ square images can be represented while $2^n \times 2^n$ square images can be represented in the other models.

The time complexity of preparing procedure for the $2^n \times 2^n$ images with 4-channels and $2^q$ color tones representation models is shown in Table 2.

It is seen from Table 2 that the QRMW model shows a quadratic decrease when it is compared with the MCQI model and linear decrease when it is compared with the QMCR model.

## 3.2. Image retrieving procedure

It is an important process to retrieving the image from a quantum state, efficiently. Quantum measurement is a unique way to recover classical information from a quantum state. The following operations are performed to obtain pixel values from the $2^n \times 2^m$ QRMW image with $2^b$ wavelength channel and $2^q$ color scale in the channels.

The quantum measurement preparation operator $\Gamma_{\lambda yx}$ which will be used for the qubit sequence at the desired position and wavelength is defined as follows.

$$\Gamma_{\lambda yx} = \sqrt{2^{b+n+m}}\left(I^{\otimes q} \otimes |\lambda yx\rangle\langle\lambda yx|\right) \quad (11)$$

$\Gamma_{\lambda yx}$ sets the amplitude probability of only the corresponding $(\lambda, y, x)$ pixel to 1 and the others to 0. If we apply Eq. (11) to Eq. (10), we get

$$|P_{\lambda yx}\rangle = |f(\lambda, y, x)\rangle|\lambda yx\rangle \quad (12)$$

The state in Eq. (12) shows image information at desired channel of desired pixel. The first register of the state $|P_{\lambda yx}\rangle$ is measured at the computational basis $|0\rangle$ and $|1\rangle$ to obtain the binary sequence $f(\lambda, y, x)$ of the color information. So the color value of the corresponding $(\lambda, y, x)$ pixel can be retrieved. When these operations are repeated for each channel of pixels at each position, the classic image is retrieved from the QRMW image model. Let us compare the retrieving of our QRMW model with QMCR and MCQI models.

The color values of the pixels are stored at the basis states of qubit sequence in the QRMW and QMCR models. The color values of the pixels are stored at the probability amplitude in the MCQI model. The color values are exactly obtained in QRMW and QMCR models. The color values are approximately obtained in the MCQI model. The color values of the pixels need to be exact for the details of image and detailed operations.

## 4. Quantum image compression and color operations

In this section, we propose a compression method to reduce the number of gates used during the preparation of the QRMW image. So, the time complexity in QRMW model will be reduced. Also some color operations will be mentioned on the prepared QRMW image.

### 4.1. Quantum image compression

Quantum image Compression (QIC) is a process that reduces the amount of computational resources used in image preparation and retrieving operations such as in classical image compression techniques. The main resources of quantum computation are the basic quantum gates and operators. Also this is expressed as time complexity.

From the image preparation procedure, it is clear that the time complexity is directly related to the image size. When the image size is too large, the spending time of preparation procedure takes long time. For QIC, firstly the pixels with the same color value in different positions and/or channels in the image should be grouped. Applying operations to groups with the same information instead of performing individual operations on all the pixels will reduce the resources (proportional to the number of groups). For this purposes, Boolean expression should be obtained for the state. Methods such as Karnaugh maps, Espresso algorithm, etc. [10] can be used to minimize Boolean expressions. In this paper, Espresso algorithm is used to minimize Boolean expressions.

In our model, the operator $U_{\lambda_{yx}}$ is the process of setting the color value to the corresponding channel of the pixel in the corresponding position during the QRMW image preparation procedure. Thus the runtime is directly related to the number of application of this operator.

$U_{\lambda_{yx}}$ and $\Omega_{\lambda_{yx}}$ operators have been updated to work as a single operator for the {y} list with the same values in groups according to the minimized boolean expressions.

$$U_{\lambda_{\{y\}x}} = I^{\otimes q} \otimes \sum_{k=0}^{2^b-1} \sum_{j=0}^{2^n-1} \sum_{i=0}^{2^m-1} |kji\rangle\langle kji| + \Omega_{\lambda_{\{y\}x}} \quad (13)$$

Where $kji \neq \lambda\{y\}x$ and $\{y\} = \{y_0, y_1, \ldots, y_{n-1}\}$

$$\Omega_{\lambda_{\{y\}x}} = \sum_{i=0}^{n-1} |f(\lambda, y_0, x)\rangle|\lambda y_i x\rangle\langle 0|^{\otimes q}\langle \lambda y_i x| \quad (14)$$

Where $f(\lambda, y_0, x) = f(\lambda, y_1, x) = \cdots = f(\lambda, y_{n-1}, x)$

Let us consider a 8×4 sample image with 4-channels and different color value to illustrate the compression process by grouping on the pixels of the same color value in the QRMW image.

According to the sample of Figure 3, {y} = {0,1,2,3,4,5,6,7}. We need to use only one operator instead of eight operators for this minimization. This is the same for the other eight groups.

9-$U_{\lambda_{yx}}$ operators are used instead of 72 for the groups given in Figure 4. If we calculate compression ratio for example, we get

$$Compression\_Ratio = \left(1 - \frac{Ops\_After\_Compression}{Ops\_Before\_Compression}\right) \times 100\% \quad (15)$$

According to the equation given in Eq. (15), the compression ratio of the QRMW is 87.5%.

Let us compare the ratio of our QRMW with the other models s for four sample images which are given in Figure 5 with the size $4 \times 4$.

It is seen from Table 3 that compression ratio of QRMW model is more than MCQI and QMCR models.

## 4.2. Quantum image color operations

In this section, the operations on the images will be presented in three categories such as Complete color operations, Partial color operations, Position operations.

### 4.2.1. Complete color operations

The $U_{CC}$ operator is defined as

$$U_{CC} = X^{\otimes q} \otimes I^{\otimes b+n+m} \tag{16}$$

The $U_{CC}$ operator applies quantum $X$ gate to each qubit of the qubit sequence in the first register in which the color information is stored for the $|I\rangle$ quantum image. This operator reverses all qubits in the color qubit sequence. So it sets the current color values to the opposite values. So, all the colors in the whole image will change. Application of the $U_{CC}$ operator to the $|I\rangle$ quantum image is given as follows

$$U_{CC}|I\rangle = \frac{1}{\sqrt{2^{b+n+m}}} \left( \sum_{\lambda=0}^{2^b-1} \sum_{y=0}^{2^n-1} \sum_{x=0}^{2^m-1} |2^q - 1 - f(\lambda, y, x)\rangle |\lambda y x\rangle \right) \tag{17}$$

If we apply Eq. (17) to sample image Figure 6b, we get image Figure 6c. The quantum circuit of the $U_{CC}$ operator and the changes in the sample image are shown in Figure 6. The $U_{CC}$ operator can be expanded for combinations of different numbers and positions of $q$-qubits in the first register in which color information is stored. For example, $U_{CC} = X \otimes I^{\otimes \alpha} \otimes X \otimes I^{\otimes q-\alpha-2} \otimes I^{\otimes b+n+m}$, where $\alpha$ is desired arbitrary number for first register. $\alpha$ can be maximum $q - 2$.

### 4.2.2. Partial color operations

The channel-controlled $U_{(PC)_k}$ operator which is prepared to process colors in a particular channel is given as follows,

$$U_{(PC)_k} = (b - CNOT)_k^{\otimes q} \otimes I^{\otimes b+n+m} \tag{18}$$

Where b: channel qubit number. For the $|I\rangle$ quantum image, the $U_{(PC)_k}$ operator applies *b-CNOT* gate on the desired qubits of the qubit sequence in the first register in which color information is stored. This operator reverses the qubits in the qubit sequence in which the color information of the corresponding channel (*k*) is stored. So, the color of the corresponding channel in the whole image will change only. Application of the $U_{(PC)_k}$ operator to the $|I\rangle$ quantum image is given as follows.

$$U_{(PC)_k}|I\rangle = \frac{1}{\sqrt{2^{b+n+m}}} \left( \sum_{\lambda=0,\lambda \neq k}^{2^b-1} \sum_{y=0}^{2^n-1} \sum_{x=0}^{2^m-1} |f(\lambda,y,x)\rangle|\lambda yx\rangle + |2^q - 1 - f(k,y,x)\rangle|kyx\rangle \right) \quad (19)$$

The quantum circuit of the $U_{(PC)_k}$ operator and the changes in the sample image are shown in Figure 7. The $U_{(PC)_k}$ operator can be expanded for combinations of different numbers and positions of *q*-qubits in the first register in which color information is stored. For example, $U_{(PC)_k} = (b - CNOT)_k \otimes I_k^{\otimes \alpha} \otimes (b - CNOT)_k \otimes I_k^{\otimes q-\alpha-2} \otimes I^{\otimes b+n+m}$, where $\alpha$ is desired arbitrary number for first register. $\alpha$ can be maximum $q - 2$.

The $U_{CH}$ and $U_{(CH)_k}$ operators which are prepared for changing the color values of all or some of the channels in the quantum image are given as follows

$$U_{CH} = I^{\otimes q} \otimes X^{\otimes b} \otimes I^{\otimes n+m} \quad (20)$$

$$U_{(CH)_k} = I^{\otimes q} \otimes (k - CNOT)^{\otimes b} \otimes I^{\otimes n+m} \quad (21)$$

For the $|I\rangle$ quantum image, the $U_{CH}$ operator applies the *X* gate to all the qubits in the qubit array in the second register where the channel information is stored. The operator $U_{CH}$ changes the colors of all channels. The operator $U_{(CH)_k}$ applies *k-CNOT* gate to the desired qubits in the qubit sequence in the second register where the channel information is stored. $U_{(CH)_k}$ changes the colors of desired channels. Application of the $U_{CH}$ operator to the $|I\rangle$ quantum image is given as follows.

$$U_{CH}|I\rangle = \frac{1}{\sqrt{2^{b+n+m}}} \left( \sum_{\lambda=0}^{2^b-1} \sum_{y=0}^{2^n-1} \sum_{x=0}^{2^m-1} |f(\lambda,y,x)\rangle |2^b - 1 - \lambda\rangle |yx\rangle \right) \quad (22)$$

The quantum circuit of the $U_{CH}$ operator and the changes in the sample image are shown in Figure 8. The States of the red, green, blue and alpha channels are respectively are $|00\rangle$, $|01\rangle$, $|10\rangle$, $|11\rangle$. The $U_{(CH)_k}$ operator is the extended version of the $U_{CH}$ operator for combinations of different numbers and positions of the $b$-qubits in the second register in which channel information is stored.

### 4.2.3. Position operations

The operators $U_{PO}$ and $U_{(PO)_{yx}}$ for position operations are shown as

$$U_{PO} = I^{\otimes q+b} \otimes X^{\otimes n+m} \quad (23)$$

$$U_{(PO)_{yx}} = I^{\otimes q+b} \otimes (yx - CNOT)^{\otimes n+m} \quad (24)$$

For the $|I\rangle$ quantum image, the $U_{PO}$ operator applies the $X$ gate to all qubits of the qubit sequence in the third register where the position information is stored and the $U_{(PO)_{yx}}$ operator applies $yx$-CNOT gate to the corresponding qubits in the third register. The $U_{PO}$ operator reverses the qubits in the qubit sequence in which the position information is stored. So, the all of pixels has been relocated in the image. $U_{(PO)_{yx}}$ will change only the specified row/column pixels, not all the pixels. Application of the $U_{PO}$ operator to the $|I\rangle$ quantum image is given as follows.

$$U_{PO}|I\rangle = \frac{1}{\sqrt{2^{b+n+m}}} \left( \sum_{\lambda=0}^{2^b-1} \sum_{y=0}^{2^n-1} \sum_{x=0}^{2^m-1} |f(\lambda,y,x)\rangle |\lambda\rangle |2^b - 1 - (yx)\rangle \right) \quad (25)$$

The quantum circuit of the $U_{PO}$ operator and the changes in the sample image are shown in Figure 9. The $U_{(PO)_{yx}}$ operator is the extended version of the $U_{PO}$ operator for combinations of different numbers and positions of the ($n+m$)-qubits in the third register in which position information is stored.

**5. Conclusion**

Images with different wavelengths of light are now used in many fields, for example, satellite images. Quantum-based image processing is an effective and important approach to reduce the high real-time computing requirements of the classical image processing. In this respect, quantum representation of image is an important step.

In this paper, we present QRMW model with multi wavelength and some processing on the image for quantum representation. QRMW has been compared with the existing quantum representation models of image. Also, advantages/disadvantages of our model have been presented. Also model has been simulated at MS-Liquid which is quantum programming language developed by Microsoft [11].

An important issue is the speed of data processing, as well as data transfer and processing in any information processing. So, the Quantum Representation of Multi Wavelength Images (QRMW) model offers faster processing performance with fewer qubit sources than other quantum representation models.

**Acknowledgements:** We would like to thank to Referees for their valuable suggestions. This paper was produced from the part of the PhD study of Engin ŞAHİN.

**Tables**

| Model | Qubits | Arbitrary qubits | Total qubits |
|---|---|---|---|
| **QRMW** | q + 2 + 2n | - | q + 2 + 2n |
| **MCQI** | 3 + 2n | - | 3 + 2n |
| **QMCR** | 4q + 2n | 4q | 8q+2n |

**Table 1.** Numbers of qubits of the QRMW, MCQI and QMCR with 4-channels and $2^q$ color scale for $2^n \times 2^n$ images.

| Model | 1st Step | 2nd Step |
|---|---|---|
| **QRMW** | $O(q+2+2n)$ | $O(q \times 2^{(2+2n)})$ |
| **MCQI** | $O(3+2n)$ | $O(2^{(3+4n)})$ |
| **QMCR** | $O(4q+2n)$ | $O(qn \times 2^{(2+2n)})$ |

**Table 2.** The time complexity information for preparing procedure of QRMW, MCQI and QMCR images.

| Model | (a) | (b) | (c) | (d) | AVG |
|---|---|---|---|---|---|
| **QRMW** | 91.67% | 83.34% | 97.92% | 0% | 68.23% |
|  | 4/48 | 8/48 | 1/48 | 48/48 |  |
| **MCQI** | 75% | 50% | 93.75% | 0% | 54.69% |
|  | 512/2048 | 1024/2048 | 128/2048 | 2048/2048 |  |
| **QMCR** | 75% | 50% | 93.75% | 0% | 54.69% |
|  | 96/384 | 192/384 | 24/384 | 384/384 |  |

**Table 3.** Compression ratios of QRMW, MCQI and QMCR models for **(a)-(d)** sample images in Figure 5.

**Figures**

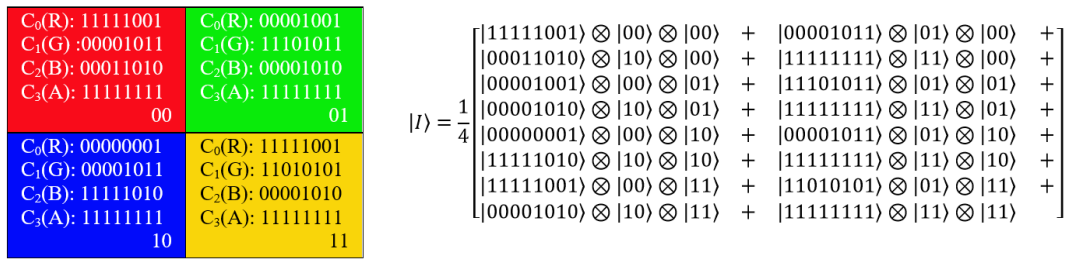

**Figure 1.** 2×2 image sample and expression of QRMW model.

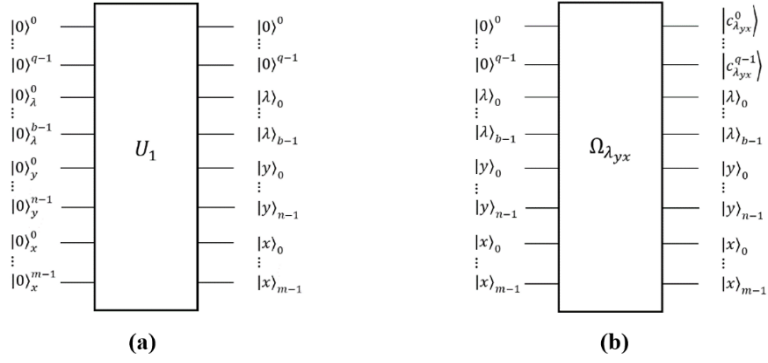

**Figure 2. (a)** The Black-Box circuit of the first step of the preparation procedure, **(b)** the Black-Box circuit of the second step which is the process of setting color to the corresponding channel of the corresponding position.

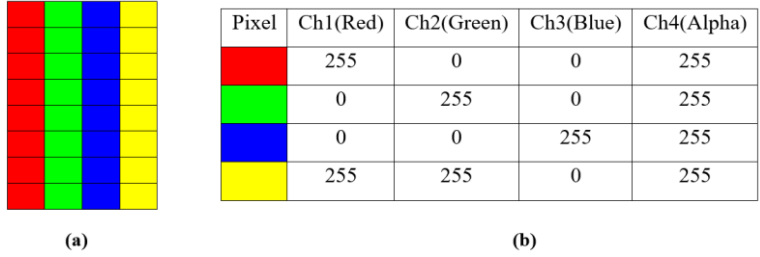

**Figure 3. (a)** 8×4 sample image with 4 channels and different color values, **(b)** channel and color information of pixels.

| Color | Channel | Y | X |
|---|---|---|---|
| 11111111 | 00 | 000 | 00 |
| 11111111 | 00 | 001 | 00 |
| 11111111 | 00 | 010 | 00 |
| 11111111 | 00 | 011 | 00 |
| 11111111 | 00 | 100 | 00 |
| 11111111 | 00 | 101 | 00 |
| 11111111 | 00 | 110 | 00 |
| 11111111 | 00 | 111 | 00 |
| Group 1 (red) | | | |

| Color | Channel | Y | X |
|---|---|---|---|
| 11111111 | 01 | 000 | 01 |
| 11111111 | 01 | 001 | 01 |
| 11111111 | 01 | 010 | 01 |
| 11111111 | 01 | 011 | 01 |
| 11111111 | 01 | 100 | 01 |
| 11111111 | 01 | 101 | 01 |
| 11111111 | 01 | 110 | 01 |
| 11111111 | 01 | 111 | 01 |
| Group 2 (green) | | | |

| Color | Channel | Y | X |
|---|---|---|---|
| 11111111 | 10 | 000 | 10 |
| 11111111 | 10 | 001 | 10 |
| 11111111 | 10 | 010 | 10 |
| 11111111 | 10 | 011 | 10 |
| 11111111 | 10 | 100 | 10 |
| 11111111 | 10 | 101 | 10 |
| 11111111 | 10 | 110 | 10 |
| 11111111 | 10 | 111 | 10 |
| Group 3 (blue) | | | |

| Color | Channel | Y | X |
|---|---|---|---|
| 11111111 | 00 | 000 | 11 |
| 11111111 | 00 | 001 | 11 |
| 11111111 | 00 | 010 | 11 |
| 11111111 | 00 | 011 | 11 |
| 11111111 | 00 | 100 | 11 |
| 11111111 | 00 | 101 | 11 |
| 11111111 | 00 | 110 | 11 |
| 11111111 | 00 | 111 | 11 |
| Group 4 (red2) | | | |

| Color | Channel | Y | X |
|---|---|---|---|
| 11111111 | 01 | 000 | 11 |
| 11111111 | 01 | 001 | 11 |
| 11111111 | 01 | 010 | 11 |
| 11111111 | 01 | 011 | 11 |
| 11111111 | 01 | 100 | 11 |
| 11111111 | 01 | 101 | 11 |
| 11111111 | 01 | 110 | 11 |
| 11111111 | 01 | 111 | 11 |
| Group 5 (green2) | | | |

| Color | Channel | Y | X |
|---|---|---|---|
| 11111111 | 11 | 000 | 00 |
| 11111111 | 11 | 001 | 00 |
| 11111111 | 11 | 010 | 00 |
| 11111111 | 11 | 011 | 00 |
| 11111111 | 11 | 100 | 00 |
| 11111111 | 11 | 101 | 00 |
| 11111111 | 11 | 110 | 00 |
| 11111111 | 11 | 111 | 00 |
| Group 6 (alpha1) | | | |

| Color | Channel | Y | X |
|---|---|---|---|
| 11111111 | 11 | 000 | 01 |
| 11111111 | 11 | 001 | 01 |
| 11111111 | 11 | 010 | 01 |
| 11111111 | 11 | 011 | 01 |
| 11111111 | 11 | 100 | 01 |
| 11111111 | 11 | 101 | 01 |
| 11111111 | 11 | 110 | 01 |
| 11111111 | 11 | 111 | 01 |
| Group 7 (alpha2) | | | |

| Color | Channel | Y | X |
|---|---|---|---|
| 11111111 | 11 | 000 | 10 |
| 11111111 | 11 | 001 | 10 |
| 11111111 | 11 | 010 | 10 |
| 11111111 | 11 | 011 | 10 |
| 11111111 | 11 | 100 | 10 |
| 11111111 | 11 | 101 | 10 |
| 11111111 | 11 | 110 | 10 |
| 11111111 | 11 | 111 | 10 |
| Group 8 (alpha2) | | | |

| Color | Channel | Y | X |
|---|---|---|---|
| 11111111 | 11 | 000 | 11 |
| 11111111 | 11 | 001 | 11 |
| 11111111 | 11 | 010 | 11 |
| 11111111 | 11 | 011 | 11 |
| 11111111 | 11 | 100 | 11 |
| 11111111 | 11 | 101 | 11 |
| 11111111 | 11 | 110 | 11 |
| 11111111 | 11 | 111 | 11 |
| Gorup 9 (alpha4) | | | |

**Figure 4.** The nine groups in positions with the same color information before quantum image preparation.

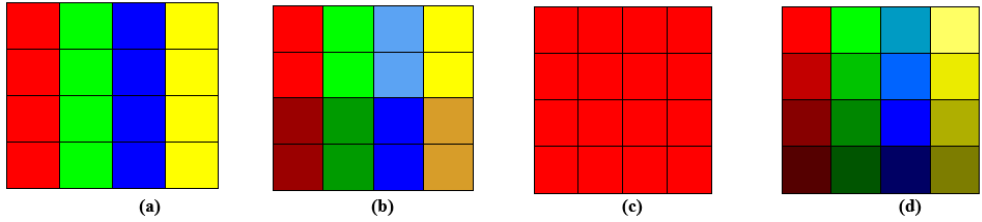

**Figure 5.** 4 × 4 Sample images with 256 color scale and 3-channels.

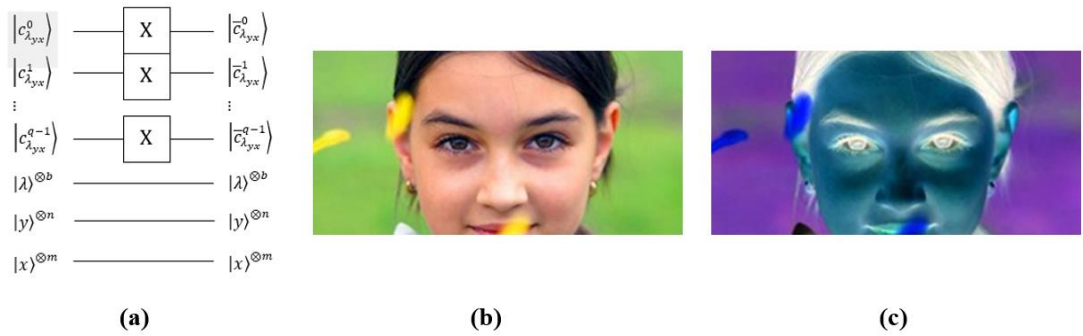

**Figure 6. (a)** Quantum circuit of $U_{CC}$ operator for QRMW model **(b)** sample image with 3-channels **(c)** the image after $U_{CC}$ is applied.

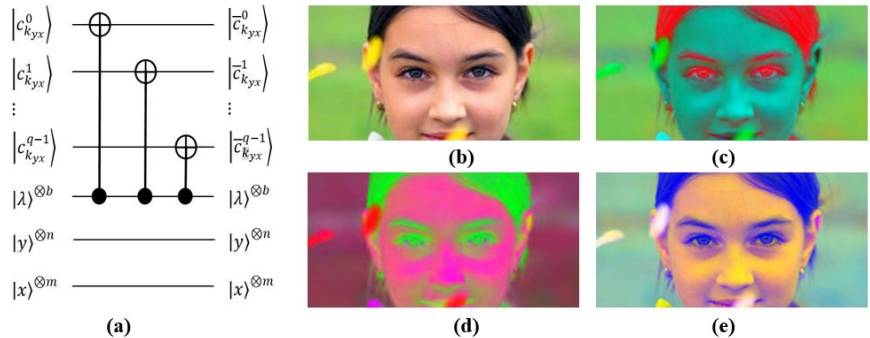

**Figure 7. (a)** Quantum circuit of $U_{(PC)_k}$ operator for QRMW model **(b)** sample image with 3-channels **(c)** the image after $U_{(PC)_0}$ is applied **(d)** the image after $U_{(PC)_1}$ is applied **(e)** the image after $U_{(PC)_2}$ is applied.

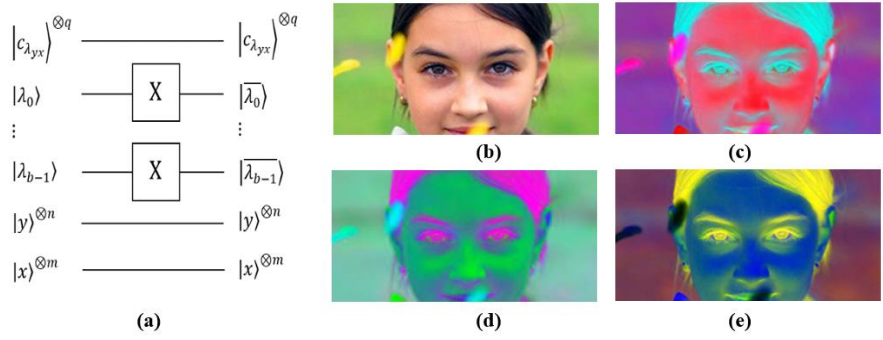

**Figure 8.** **(a)** Quantum circuit of $U_{CH}$ operator for QRMW model **(b)** sample image with 4-channels **(c)** the image after $U_{CH}$ is applied **(d)** the image after $U_{(CH)_1}$ is applied **(e)** the image after $U_{(CH)_2}$ is applied

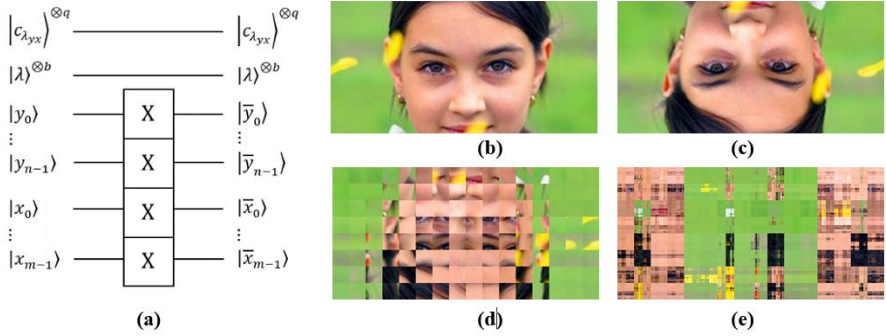

**Figure 9.** **(a)** Quantum circuit of $U_{PO}$ operator for QRMW model **(b)** sample image **(c)** the image after $U_{PO}$ is applied **(d)** the image after $U_{(PO)_{\{1,2,3\}\{7,8,9\}}}$ is applied **(e)** the image after $U_{(PO)_{\{1,3,5\}\{8,10,12\}}}$ is applied